\documentstyle[12pt,epsfig]{article}

\catcode`\@=11
\long\def\@makefntext#1{
\protect\noindent \hbox to 3.2pt {\hskip-.9pt  
$^{{\ninerm\@thefnmark}}$\hfil}#1\hfill}		

\def\@makefnmark{\hbox to 0pt{$^{\@thefnmark}$\hss}}  
	
\def\ps@myheadings{\let\@mkboth\@gobbletwo
\def\@oddhead{\hbox{}
\rightmark\hfil\ninerm\thepage}   
\def\@oddfoot{}\def\@evenhead{\ninerm\thepage\hfil
\leftmark\hbox{}}\def\@evenfoot{}
\def\sectionmark##1{}\def\subsectionmark##1{}}

\renewcommand{\thefootnote}{\fnsymbol{footnote}}

\newcounter{sectionc}\newcounter{subsectionc}\newcounter{subsubsectionc}
\renewcommand{\section}[1] {\vspace*{0.6cm}\addtocounter{sectionc}{1} 
\setcounter{subsectionc}{0}\setcounter{subsubsectionc}{0}\noindent 
	{\normalsize\bf\thesectionc. #1}\par\vspace*{0.4cm}}
\renewcommand{\subsection}[1] {\vspace*{0.6cm}\addtocounter{subsectionc}{1} 
	\setcounter{subsubsectionc}{0}\noindent 
	{\normalsize\it\thesectionc.\thesubsectionc. #1}\par\vspace*{0.4cm}}
\renewcommand{\subsubsection}[1]
{\vspace*{0.6cm}\addtocounter{subsubsectionc}{1}
	\noindent {\normalsize\rm\thesectionc.\thesubsectionc.\thesubsubsectionc. 
	#1}\par\vspace*{0.4cm}}

\newcounter{appendixc}
\newcounter{subappendixc}[appendixc]
\newcounter{subsubappendixc}[subappendixc]

\renewcommand{\appendix}[1] {\vspace*{0.6cm}
        \refstepcounter{appendixc}
        \setcounter{figure}{0}
        \setcounter{table}{0}
        \setcounter{equation}{0}
        \renewcommand{\thefigure}{\Alph{appendixc}.\arabic{figure}}
        \renewcommand{\thetable}{\Alph{appendixc}.\arabic{table}}
        \renewcommand{\theappendixc}{\Alph{appendixc}}
        \renewcommand{\theequation}{\Alph{appendixc}.\arabic{equation}}
        \noindent{\bf Appendix \theappendixc #1}\par\vspace*{0.4cm}}

\def\abstracts#1{{
	\centering{\begin{minipage}{12.2truecm}\footnotesize\baselineskip=12pt\noindent
	\centerline{\footnotesize ABSTRACT}\vspace*{0.3cm}
	\parindent=0pt #1
	\end{minipage}}\par}} 

\renewenvironment{thebibliography}[1]
	{\begin{list}{\arabic{enumi}.}
	{\usecounter{enumi}\setlength{\parsep}{0pt}
\setlength{\leftmargin 1.25cm}{\rightmargin 0pt}
	 \setlength{\itemsep}{0pt} \settowidth
	{\labelwidth}{#1.}\sloppy}}{\end{list}}

\newcounter{itemlistc}
\newcounter{romanlistc}
\newcounter{alphlistc}
\newcounter{arabiclistc}

\newcommand{\fcaption}[1]{
        \refstepcounter{figure}
        \setbox\@tempboxa = \hbox{\footnotesize Fig.~\thefigure. #1}
        \ifdim \wd\@tempboxa > 6in
           {\begin{center}
        \parbox{6in}{\footnotesize\baselineskip=12pt Fig.~\thefigure. #1}
            \end{center}}
        \else
             {\begin{center}
             {\footnotesize Fig.~\thefigure. #1}
              \end{center}}
        \fi}

\newcommand{\tcaption}[1]{
        \refstepcounter{table}
        \setbox\@tempboxa = \hbox{\footnotesize Table~\thetable. #1}
        \ifdim \wd\@tempboxa > 6in
           {\begin{center}
        \parbox{6in}{\footnotesize\baselineskip=12pt Table~\thetable. #1}
            \end{center}}
        \else
             {\begin{center}
             {\footnotesize Table~\thetable. #1}
              \end{center}}
        \fi}

\def\@citex[#1]#2{\if@filesw\immediate\write\@auxout
	{\string\citation{#2}}\fi
\def\@citea{}\@cite{\@for\@citeb:=#2\do
	{\@citea\def\@citea{,}\@ifundefined
	{b@\@citeb}{{\bf ?}\@warning
	{Citation `\@citeb' on page \thepage \space undefined}}
	{\csname b@\@citeb\endcsname}}}{#1}}

\newif\if@cghi
\def\cite{\@cghitrue\@ifnextchar [{\@tempswatrue
	\@citex}{\@tempswafalse\@citex[]}}
\def\citelow{\@cghifalse\@ifnextchar [{\@tempswatrue
	\@citex}{\@tempswafalse\@citex[]}}
\def\@cite#1#2{{$\null^{#1}$\if@tempswa\typeout
	{IJCGA warning: optional citation argument 
	ignored: `#2'} \fi}}

 1
 1
 1

\font\ninerm=cmr9

\newcommand{\n}{\hspace*{-2.5mm}}
\newcommand{\gsim}{\;\rlap{\lower 3.5 pt \hbox{$\mathchar \sim$}} \raise 1pt
 \hbox {$>$}\;}
\newcommand{\lsim}{\;\rlap{\lower 3.5 pt \hbox{$\mathchar \sim$}} \raise 1pt
 \hbox {$<$}\;}

\begin{document}
\vskip-2.cm
\centerline{\normalsize\hfill MPI/PhT/97--056}
\centerline{\normalsize\hfill hep--ph/9709261}
\centerline{\normalsize\hfill September 1997}
\vskip.5cm
\centerline{\normalsize\bf 
THEORETICAL ASPECTS OF HADRON PHOTOPRODUCTION\footnote{To appear in the
{\it Proceedings of the Ringberg Workshop: New Trends in HERA Physics},
Ringberg Castle, Germany, 25--30 May 1997.}
}
\vspace*{0.6cm}
\centerline{\footnotesize BERND A. KNIEHL}
\baselineskip=13pt
\centerline{\footnotesize\it
Max-Planck-Institut f\"ur Physik (Werner-Heisenberg-Institut)}
\baselineskip=12pt
\centerline{\footnotesize\it 
F\"ohringer Ring 6, 80805 Munich, Germany}
\centerline{\footnotesize E-mail: kniehl@vms.mppmu.mpg.de}
\vspace*{0.9cm}
\abstracts{
We review recent developments in the theoretical description of inclusive
single-hadron production at next-to-leading order in the parton model of
quantum chromodynamics.
Fragmentation functions are extracted from fits to data of inclusive pion and 
kaon production in $e^+e^-$ annihilation at different centre-of-mass energies.
Exploiting sensitivity to the scaling violation, one can simultaneously fit
the asymptotic scale parameter $\Lambda$ so as to obtain an independent
determination of the strong coupling constant $\alpha_s$.
Owing to the factorization theorem, the fragmentation functions only depend on 
the species of the produced particles, but not on the process by which they 
are produced.
This allows one to make absolute theoretical predictions for inclusive pion
and kaon production in other types of experiments such as photon-photon,
photon-hadron, or hadron-hadron scattering.
Recent data of photoproduction taken by the H1 and ZEUS Collaborations at DESY
HERA nicely agree with such next-to-leading-order predictions.
}
\normalsize\baselineskip=15pt
\setcounter{footnote}{0}
\renewcommand{\thefootnote}{\alph{footnote}}

\section{Introduction}

The Lagrangian of quantum chromodynamics (QCD) contains quarks and gluons as
elementary fields.
Allowing for these particles to appear as asymptotic states, we can evaluate
scattering amplitudes perturbatively, in principle with arbitrary precision.
Of course, this picture needs to be complemented by the principle of
confinement of colour;
experiments detect hadrons rather than quarks and gluons.
Nevertheless, this simplified computational procedure is very successful
in describing the production of jets of hadrons at high centre-of-mass (c.m.)
energies ($\sqrt s\,$) in $e^+e^-$ annihilation and at high transverse momenta
($p_T$) in scattering processes.
Due to parton-hadron duality, clustering partons in the final state according
to certain jet definitions yields a useful approximation, although this does
not account for any details of hadronization.

On the other hand, experiments are providing us with copious information on 
the inclusive production of single hadrons, which cannot be interpreted along
these lines.
In this case, we need a detailed concept for describing how partons turn into
hadrons.
In the framework of the QCD-improved parton model, this is achieved by
introducing fragmentation functions (FF's), $D_a^h(x,\mu^2)$.
The value of $D_a^h(x,\mu^2)$ corresponds to the probability for the parton
$a$ produced at short distance $1/\mu$ to form a jet that includes the hadron
$h$ carrying the fraction $x$ of the longitudinal momentum of $a$.
Unfortunately, it is not yet understood how the FF's can be derived from first
principles, in particular for hadrons with masses smaller than or comparable
to the asymptotic scale parameter, $\Lambda$.
However, given their $x$ dependence at some scale $\mu$, the evolution with
$\mu$ may be computed perturbatively in QCD using the 
Altarelli-Parisi\cite{gri} (AP) equations.
This allows us to test QCD quantitatively within one experiment observing
single hadrons at different values of $\sqrt s$ (in the case of $e^+e^-$
annihilation) of $p_T$ (in the case of scattering).
Moreover, the factorization theorem guarantees that the $D_a^h$ functions are
independent of the process in which they have been determined, and represent a
universal property of $h$.
This enables us to make quantitative predictions for other types of
experiments as well.
To summarize, having extracted FF's from fits to experimental data, we may
test their $\mu$ dependence as predicted by the AP equations and their
universality as postulated by the factorization theorem.

After the pioneering leading-order (LO) analyses of pion, kaon,\cite{bar} and
charmed-meson\cite{seh} FF's in the late 70's, there had long been no 
progress on the theoretical side of this field.
This may partly be attributed to the advent of general-purpose Monte Carlo
(MC) event generators based on LO parton-level matrix elements in the early
80's, which were soon to become very popular in the experimental community. 
In these computer programs, the fragmentation into hadrons is simulated 
according to certain phenomenological model assumptions, {\it e.g.}, the
cluster algorithm in the case of HERWIG or the LUND string model in the case
of PYTHIA.
Although such MC packages often lead to satisfactory descriptions of the data,
from the theoretical point of view, their drawback is that this happens at the
expense of introducing a number of ad-hoc fine-tuning parameters, which do not
originate in the QCD Lagrangian.
Furthermore, in the MC approach, it seems impossible to implement the
factorization of final-state collinear singularities, which impedes a
consistent extension to next-to-leading order (NLO).
On the contrary, in the QCD-improved parton model, such singularities are
absorbed into the bare (infinite) FF's so as to render them renormalized
(finite) in a way quite similar to the procedure for the parton density 
functions (PDF's) at the incoming legs.
Therefore, a meaningful quantitative test of QCD can only be performed in the
parton model endowed with FF's at NLO.

Some time ago, NLO FF sets for $\pi^0$,\cite{chi} $\pi^\pm$, $K^\pm$, and
$\eta$ mesons\cite{gre} have been constructed through fits to data of $e^+e^-$
annihilation generated with HERWIG.
By contrast, our procedure, and partly that of 
Ref.~4, has been
to fit to genuine experimental $e^+e^-$ data.
We have introduced LO and NLO FF sets for charged pions and
kaons\cite{bkk1,bkk2} and for neutral kaons.\cite{bkk3}
The results presented here are obtained with the up-to-date sets of
Refs.~7 and 8, which are based on data from SLAC PEP with
$\sqrt s=29$~GeV and CERN LEP1 with $\sqrt s=M_Z$.
In contrast to 
Refs.~4 and 5, 
Refs.~6--8 also
provide ready-to-use parameterizations of the $\mu$ dependence of the FF's.

This presentation is organized as follows.
In Section~2, we shall report the essential features of our fitting procedure
for charged pions and kaons\cite{bkk2} and neutral kaons.\cite{bkk3}
In Sections~3 and 4, we shall confront recent data of inclusive pion and kaon
production collected by the H1\cite{abt,adl,joh} and ZEUS
Collaborations\cite{der} in $\gamma p$ scattering at DESY HERA and by the UA1
Collaboration\cite{boc1,boc2} in $p\bar p$ scattering at the CERN S$p\bar p$S
collider, respectively, with NLO predictions based on our FF's.
Our conclusions will be summarized in Section~5.

\boldmath
\section{Fit to data of $e^+e^-$ annihilation}
\unboldmath

The most direct way to obtain information on the FF's of hadrons is to analyze
their energy spectrum measured in $e^+e^-$ annihilation, where the theoretical
predictions are not obscured by additional nonperturbative input, {\it e.g.},
in the form of PDF's for the incoming particles.
At NLO in the parton model with $n_f$ massless quark flavours, the
differential cross section of $e^+e^-\to \gamma/Z\to h+X$, normalized to the
total hadronic cross section $\sigma_{\rm tot}$, is given by
\begin{equation}
\frac{1}{\sigma_{\rm tot}}\,\frac{d\sigma}{dx}
=\sum_a\int_x^1\frac{dz}{z}D_a^h\left(\frac{x}{z},M_h^2\right)
\frac{1}{\sigma_{\rm tot}}\,\frac{d\sigma_a}{dz}\left(z,\mu^2,M_h^2\right),
\end{equation}
where $x=2E_h/\sqrt s$ is the fraction of the beam energy carried by $h$,
$a=g,q_1,\ldots,\bar q_{n_f}$, $\mu$ is the renormalization scale, and $M_h$
is the factorization scale.
The parton-level cross sections read
\begin{eqnarray}
\frac{1}{\sigma_{\rm tot}}\,\frac{d\sigma_{q_i}}{dx}\left(x,\mu^2,M_h^2\right)
&\n=\n&\frac{e_{q_i}^2}{\sum_{i=1}^{n_f}e_{q_i}^2}
\left\{\delta(1-x)+\frac{\alpha_s(\mu^2)}{2\pi}
\left[P_{q\to q}^{(0,T)}(x)\ln\frac{s}{M_h^2}+C_q(x)\right]\right\},
\nonumber\\
\frac{1}{\sigma_{\rm tot}}\,\frac{d\sigma_g}{dx}\left(x,\mu^2,M_h^2\right)
&\n=\n&2\frac{\alpha_s(\mu^2)}{2\pi}
\left[P_{q\to g}^{(0,T)}(x)\ln\frac{s}{M_h^2}+C_g(x)\right],
\label{eq:ee}
\end{eqnarray}
where $e_{q_i}$ is the effective coupling of $q_i$ to the photon and the
$Z$ boson including propagator adjustments and $C_a$ are the NLO
corrections.\cite{ell}
Here, $P_{a\to b}^{(0,T)}$ are the LO terms of the timelike $a\to b$ splitting
functions,
\begin{equation}
P_{a\to b}^{(T)}\left(x,\alpha_s(\mu^2)\right)=
\frac{\alpha_s(\mu^2)}{2\pi}P_{a\to b}^{(0,T)}(x)
+\left(\frac{\alpha_s(\mu^2)}{2\pi}\right)^2P_{a\to b}^{(1,T)}(x)
+{\cal O}\left(\alpha_s^3\right),
\end{equation}
which control the $\mu$ evolution of the FF's via the AP equations,
\begin{equation}
\frac{\mu^2d}{d\mu^2}D_a^h(x,\mu^2)=\sum_b\int_x^1\frac{dz}{z}
P_{a\to b}^{(T)}\left(\frac{x}{z},\alpha_s(\mu^2)\right)D_b^h(x,\mu^2).
\label{eq:ap}
\end{equation}
Analytic expressions for $P_{a\to b}^{(T)}$ may be found, {\it e.g.}, in
Ref.~16.
The integro-differential equation~(\ref{eq:ap}) may be solved either via the
Mellin transform technique or with brute force as it stands.
Good agreement is found between these two methods.\cite{bkk4}

It is natural to choose $\mu=M_h=\sqrt s$.
This eliminates the terms involving $P_{a\to b}^{(0,T)}$ in Eq.~(\ref{eq:ee}).
In $e^+e^-$ annihilation, gluon fragmentation occurs only at NLO and beyond.
To increase the sensitivity to $D_g^h$, one may select longitudinal
polarization,\cite{cow} so that the delta function in Eq.~(\ref{eq:ee}) does
not contribute, or concentrate on gluon-tagged three-jet events.\cite{cow,act}
Then, the respective NLO terms in Eq.~(\ref{eq:ee}) constitute the Born
approximation, and it is desirable to include the
next-to-next-to-leading-order corrections\cite{rij} in order to reduce the
scale and scheme dependences.
Unfortunately, these were not yet available when the analyses of
Refs.~6--8 were carried out.

The fitting procedure in 
Refs.~7 and 8 was as follows.
For each of the hadron channels $h=\pi^++\pi^-,K^++K^-,K_S^0+K_L^0$ and each
parton type $a=g,u,d,s,c,b$, we made the ansatz
\begin{equation}
D_a^h(x,\mu_0^2)=Nx^\alpha(1-x)^\beta,
\end{equation}
at the respective starting scale
\begin{equation}
\mu_0=\cases{\sqrt2~{\rm GeV}&if $a=g,u,d,s$,\cr
M(\eta_c)=2.98~{\rm GeV}&if $a=c$,\cr
M(\Upsilon)=9.46~{\rm GeV}&if $a=b$.\cr}
\end{equation}
According to the flavour composition of $h$, we respectively imposed
\begin{eqnarray}
D_u^{\pi^\pm}\left(x,\mu_0^2\right)&\n=\n&D_d^{\pi^\pm}\left(x,\mu_0^2\right),
\nonumber\\
D_u^{K^\pm}\left(x,\mu_0^2\right)&\n=\n&D_s^{K^\pm}\left(x,\mu_0^2\right),
\nonumber\\
D_u^{K^0}\left(x,\mu_0^2\right)&\n=\n&D_s^{K^0}\left(x,\mu_0^2\right).
\end{eqnarray}

The analysis of 
Ref.~7 was based on charged-pion,
charged-kaon,\cite{aih,ake} and unidentified charged-hadron data.\cite{cow,act}
Apart from charged pions and kaons, mainly protons and antiprotons contribute
to the charged-hadron yield.
Inspired by 
Ref.~21, we approximated
\begin{equation}
\frac{d\sigma^{h^\pm}}{dx}=[1+f(x)]\frac{d\sigma^{\pi^\pm}}{dx}
+\frac{d\sigma^{K^\pm}}{dx},
\end{equation}
where $f(x)=0.195-1.35\,(x-0.35)^2$.
In their charged-hadron analysis, the ALEPH Collaboration\cite{cow}
distinguished between $uds$-, $c$-, and $b$-enriched samples.
Also exploiting information on identified gluon jets,\cite{cow,act}
we were thus able to treat $g,u,s,c,b\to\pi^\pm$ and $g,u,d,c,b\to K^\pm$
fragmentation separately.
Due to the large gap in $\sqrt s$ between PEP\cite{aih} and 
LEP1,\cite{cow,act,ake} we could simultaneously determine
$\Lambda_{\overline{\rm MS}}^{(5)}$.
Thus, we had a total of
$2(\pi,K)\times5(\mbox{partons})\times3(N,\alpha,\beta)
+1(\Lambda_{\overline{\rm MS}}^{(5)})=31$ independent fit parameters.
These turned out to be tightly constrained by our LO and NLO fits.
In fact, we obtained $\chi^2/\mbox{d.o.f.}$ values of $134.4/136=0.99$ and
$125.3/136=0.92$, respectively.
We found $\Lambda_{\overline{\rm MS}}^{(5)}=108$~MeV (227~MeV) at LO (NLO),
which corresponds to $\alpha_s(M_Z^2)=0.122$ (0.118).
This nicely agrees with the latest LEP value, $(0.121\pm0.003)$.\cite{cla}

In our neutral-kaon analysis,\cite{bkk3} we adopted the 
$\Lambda_{\overline{\rm MS}}^{(5)}$ values from 
Ref.~7.
Furthermore, appealing to the flavour blindness of the gluon, we assumed that
\begin{equation}
D_g^{K^0}\left(x,\mu_0^2\right)=D_g^{K^\pm}\left(x,\mu_0^2\right),
\label{eq:dg}
\end{equation}
at $\mu_0=\sqrt2$~GeV.
Thus, the number of independent fit parameters was
$4(\mbox{partons})\times3(N,\alpha,\beta)=12$.
Our combined fit to the MARK~II\cite{sch} and ALEPH\cite{dbu} neutral-kaon
samples yielded $\chi^2/\mbox{d.o.f.}=9.9/20$ at LO and 8.6/20 at NLO.

\boldmath
\section{Comparison with data of photoproduction in $ep$ scattering}
\unboldmath

According to present HERA conditions, $E_e=27.5$~GeV positrons collide with
$E_p=820$~GeV protons in the laboratory frame, so that $\sqrt s=300$~GeV is
available in the c.m.\ frame.
It has become customary to take the rapidity of hadrons travelling in the
proton direction to be positive.
The rapidities measured in the $ep$ laboratory and c.m.\ frames are related
through
\begin{equation}
y_{\rm c.m.}=y_{\rm lab}-{1\over2}\ln{E_p\over E_e}.
\end{equation}
In photoproduction, the electron or positron beam acts like a source of
quasi-real photons, with low virtualities $-Q^2$, so that HERA is effectively
operated as a $\gamma p$ collider.
The appropriate events may be discriminated from deep-inelastic-scattering
events by electron tagging or anti-tagging.
The photon flux is well approximated by the Weizs\"acker-Williams\cite{wei}
formula,
\begin{equation}
F_\gamma^e\left(x,Q_{\rm max}^2\right)
={\alpha\over2\pi}\left[{1+(1-x)^2\over x}
\ln{Q_{\rm max}^2\over Q_{\rm min}^2}
+2m_e^2x\left({1\over Q_{\rm max}^2}-{1\over Q_{\rm min}^2}\right)\right],
\end{equation}
where $x=E_\gamma/E_e$, $Q_{\rm min}^2=m_e^2x^2/(1-x)$, and
$Q_{\rm max}^2=0.01$~GeV$^2$ (0.02~GeV$^2$) for tagged events in the case of
H1 (ZEUS).
The cross section of $ep\to h+X$ emerges from the one of $\gamma p\to h+X$
by convolution with $F_\gamma^e$.
By kinematics, $x_{\rm min}\le x\le1$,
where $x_{\rm min}=p_T\exp(-y_{\rm c.m.})/[\sqrt s-p_T\exp(y_{\rm c.m.})]$.
In conformity with the H1 and ZEUS event-selection criteria, we impose
$0.3<x<0.7$ and $0.318<x<0.431$, respectively.

It is well known that $\gamma p\to h+X$ proceeds via two distinct mechanisms.
The photon can interact either directly with the partons originating from the
proton (direct photoproduction) or via its quark and gluon content (resolved
photoproduction).
Both contributions are formally of the same order in the perturbative
expansion.
Leaving aside the proton PDF's, $F_b^p$, and the FF's, $D_c^h$, which
represent common factors, the LO cross sections are of
${\cal O}(\alpha\alpha_s)$ in both cases.
In the case of the resolved mechanism, this may be understood by observing 
that the $ab\to cd$ cross sections, which are of ${\cal O}(\alpha_s^2)$, get
dressed by photon PDF's, $F_a^\gamma$, whose leading terms are of the form
$\alpha\ln(M_\gamma^2/\Lambda^2)\propto\alpha/\alpha_s$, with $M_\gamma$ being
the corresponding factorization scale.
Here, $a,b,c,d$ denote quarks and gluons.
In fact, the two mechanisms also compete with each other numerically.
Resolved photoproduction dominates at small $p_T$ and positive $y_{\rm lab}$,
while direct photoproduction wins out at large $p_T$ and negative
$y_{\rm lab}$.

LO calculations suffer from significant theoretical uncertainties connected
with the freedom in the choice of the renormalization scale, $\mu$, in
$\alpha_s$ and the factorization scales, $M_\gamma$, $M_p$, and $M_h$, in
$F_a^\gamma$, $F_b^p$, and $D_c^h$, respectively.
In order to obtain reliable predictions, it is indispensable to proceed to NLO.
Let us first consider resolved photoproduction, which is more involved.
Starting out from the well-known LO cross section of $\gamma p\to h+X$,
one needs to include the NLO corrections, $K_{ab\to c}$, to the
parton-level cross sections, $d\sigma_{ab\to c}/dt$, to substitute the
two-loop formula for $\alpha_s$, and to endow $F_a^\gamma$, $F_b^p$, and
$D_c^h$ with NLO evolution.
This leads to
\begin{eqnarray}
&\n\n\n&\frac{d^3\sigma}{dy\,d^2p_T}=\frac{1}{\pi}
\sum_{a,b,c}\int dx_\gamma dx_p{dx_h\over x_h^2}
F_a^\gamma\left(x_\gamma,M_\gamma^2\right)F_b^p\left(x_p,M_p^2\right)
D_c^h\left(x_h,M_h^2\right)
\left[\frac{d\sigma_{ab\to c}}{dt}(s,t,\mu^2)
\right.\nonumber\\
&\n\n\n&{}\times\left.
\delta\left(1+\frac{t+u}{s}\right)
+\frac{\alpha_s(\mu^2)}{2\pi}
K_{ab\to c}\left(s,t,u,\mu^2,M_\gamma^2,M_p^2,M_h^2\right)\,
\theta\left(1+\frac{t+u}{s}\right)\right],
\label{eq:res}
\end{eqnarray}
where $a,b,c=g,q_1,\ldots,\bar q_{n_f}$, $s=(p_a+p_b)^2$, $t=(p_a-p_c)^2$,
and $u=(p_b-p_c)^2$.
The parton momenta are related to the photon, proton, and hadron momenta by
$p_a=x_\gamma p_\gamma$, $p_b=x_pp_p$, and $p_c=p_h/x_h$.
The $K_{ab\to c}$ functions may be found in 
Ref.~26 for
$M_\gamma=M_p$.
This restriction was relaxed in 
Ref.~27.

The NLO cross section of direct photoproduction emerges from Eq.~(\ref{eq:res})
by substituting $F_a^\gamma(x_\gamma,M_\gamma^2)=\delta(1-x_\gamma)$,
replacing $d\sigma_{ab\to c}/dt$ and $K_{ab\to c}$ with
$d\sigma_{\gamma b\to c}/dt$ and $K_{\gamma b\to c}$, respectively,
and omitting the sum over $a$.
The $K_{\gamma b\to c}$ functions were first derived in 
Ref.~28
setting $M_\gamma=M_p=M_h$ and taking the spin-average for incoming
photons and gluons to be 1/2.
In 
Ref.~27, the scales were disentangled and the spin-average
convention was converted to the $\overline{\rm MS}$ scheme, {\it i.e.}, to be
$1/(n-2)$, with $n$ being the dimensionality of space-time.
Analytic expressions for the $K_{\gamma b\to c}$ functions are listed in
Ref.~29.

\begin{figure}[t]
\begin{tabular}{ll}
\parbox{7.3cm}{
\epsfig{file=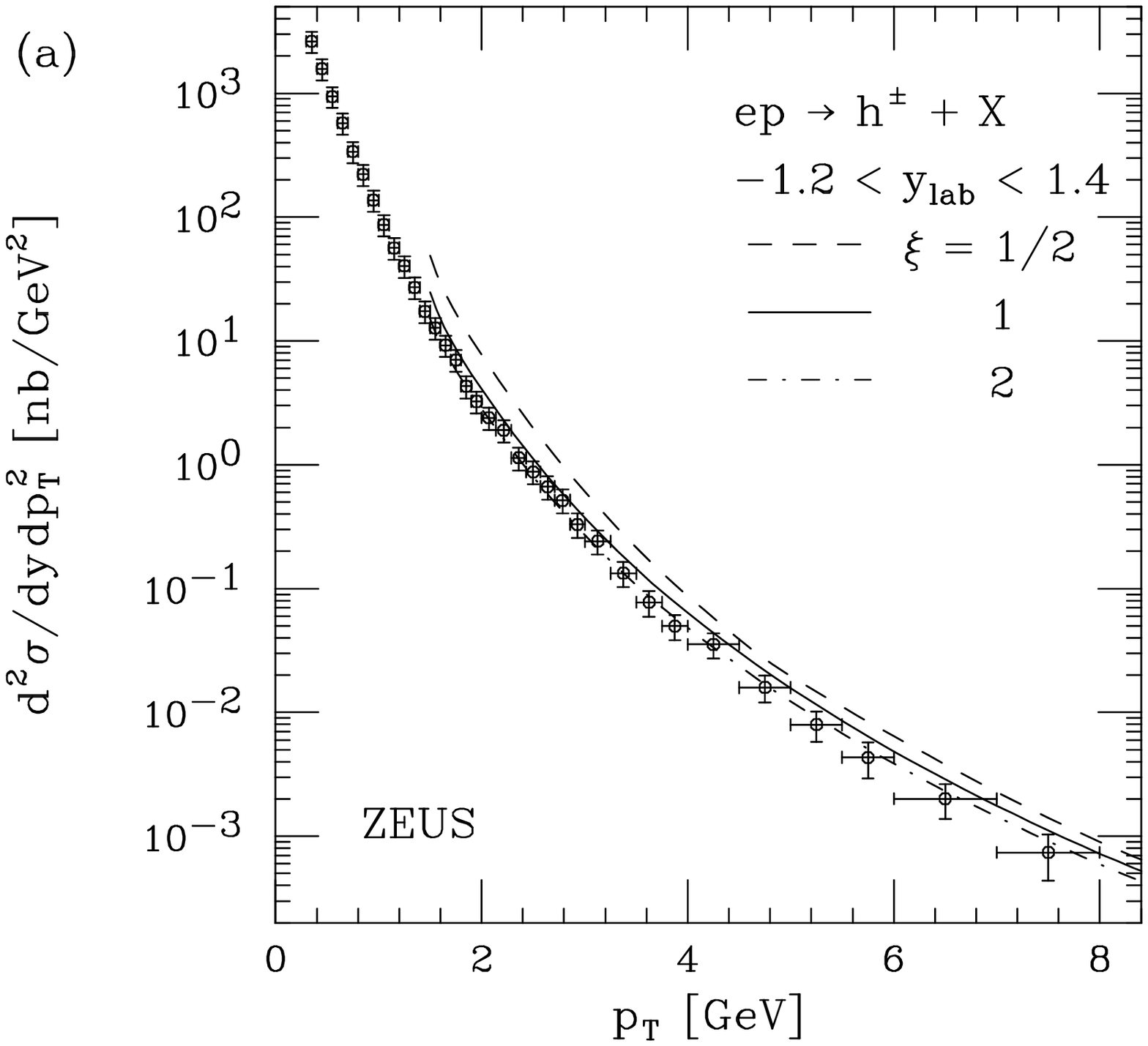,width=7.3cm}
}
&
\parbox{7.3cm}{
\epsfig{file=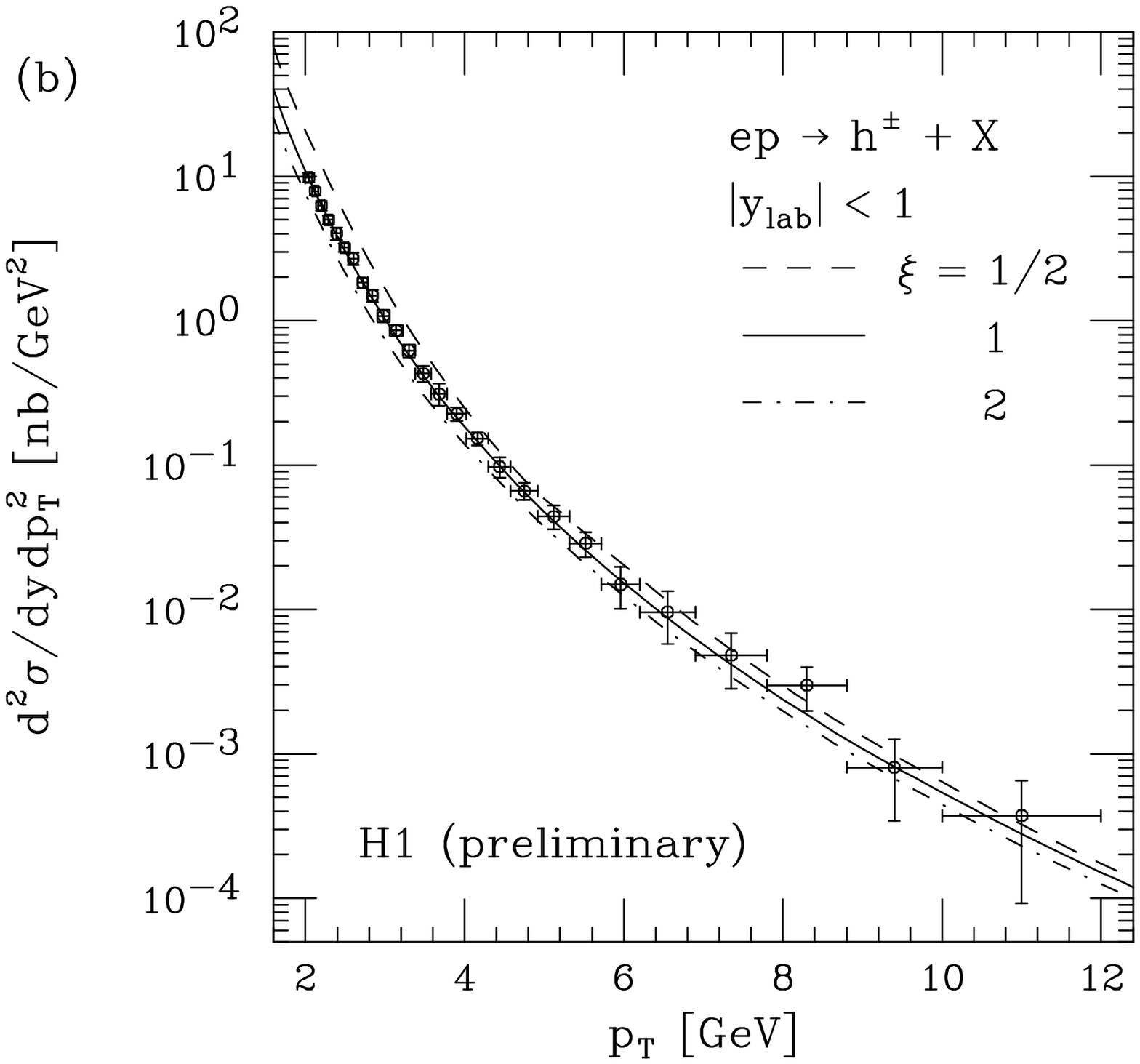,width=7.3cm}
}
\\
\parbox{7.3cm}{
\epsfig{file=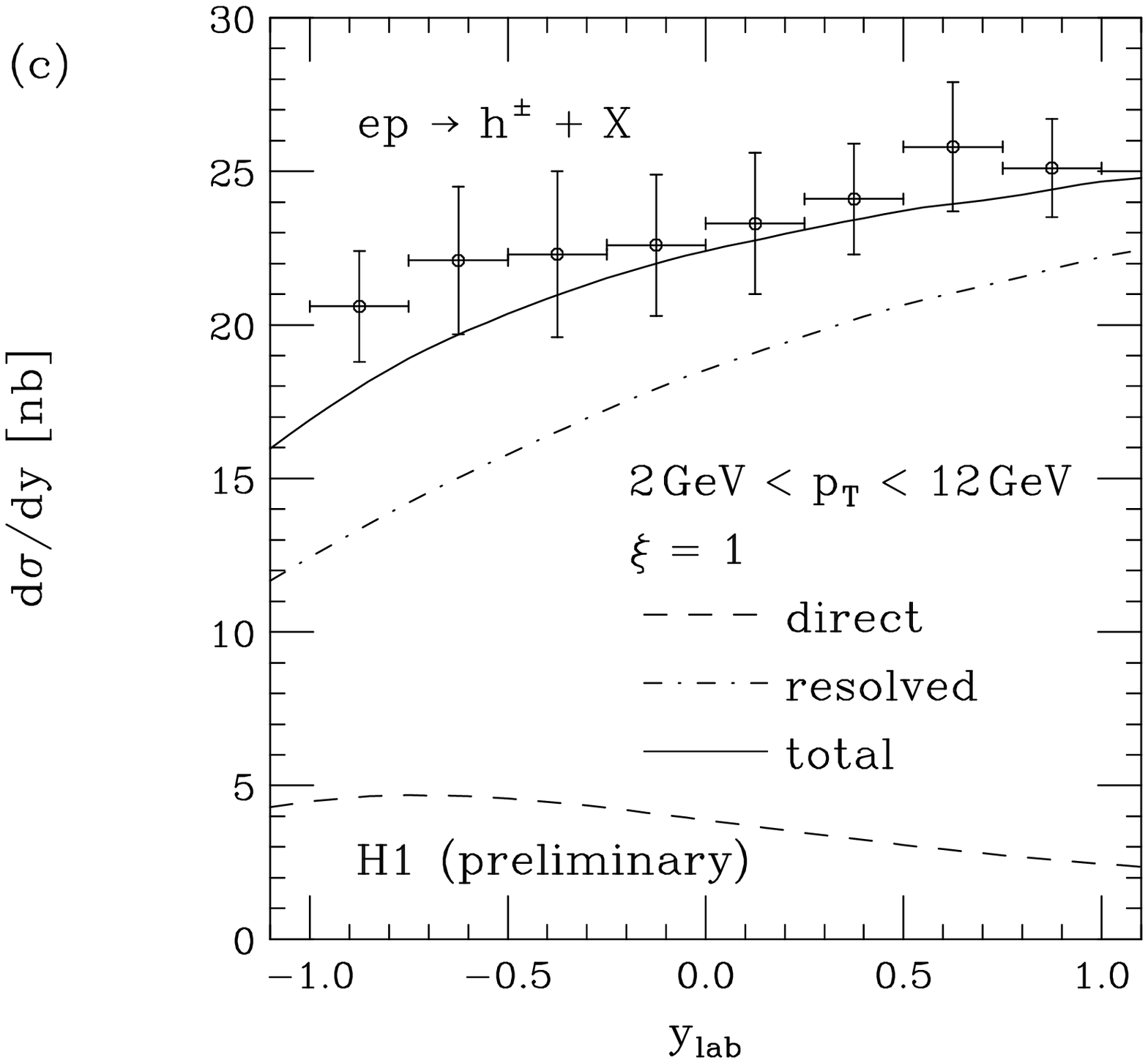,width=7.3cm}
}
&
\parbox{7.3cm}{
\epsfig{file=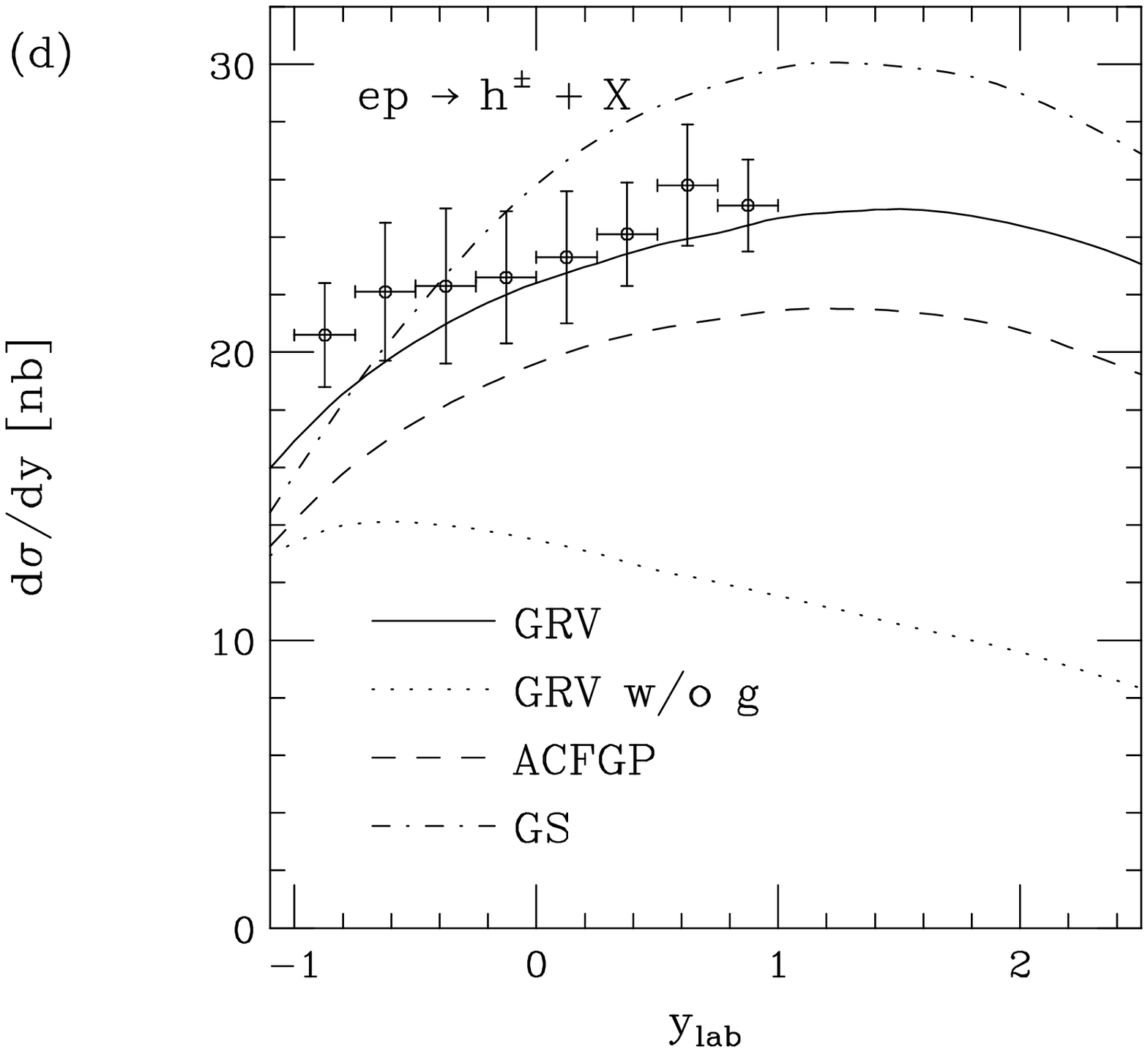,width=7.3cm}
}
\end{tabular}
\fcaption{Comparison of ZEUS\cite{der} and preliminary H1\cite{abt} data on
$ep\to h^\pm+X$ via photoproduction with NLO predictions.}
\label{fig:one}
\end{figure}

The theoretical predictions presented here are calculated at NLO in the
$\overline{\rm MS}$ scheme with $n_f=5$ quark flavours and
$\Lambda_{\overline{\rm MS}}^{(5)}=202$~MeV.
Unless otherwise stated, we use the CTEQ4M\cite{lai} proton PDF's, the
GRV\cite{grv} photon PDF's, and the BKK FF's for charged hadrons\cite{bkk2}
and neutral kaons.\cite{bkk3}
We set $\mu=M_\gamma=M_p=M_h=\xi p_T$, with $\xi=1/2,1,2$, in order to 
estimate the theoretical uncertainty.

In Fig.~\ref{fig:one}, we compare the ZEUS\cite{der} and preliminary
H1\cite{abt} data on $ep\to h^\pm+X$ via photoproduction with the
corresponding NLO predictions.
We find good agreement as for both normalization and shape.
In the upper $p_T$ range, our central prediction slightly overshoots the ZEUS
data (see Fig.~\ref{fig:one}a), while it tends to be a tiny bit below the
centres of the H1 data points, but well within their errors bars
(see Fig.~\ref{fig:one}b).
The study of the $y_{\rm lab}$ spectrum in Fig.~\ref{fig:one}c nicely
illustrates the interplay of direct and resolved photoproduction.
The direct-photon (resolved-photon) contribution peaks at negative (positive)
$y_{\rm lab}$.
At NLO, both contributions strongly depend on the factorization scheme and
scale associated with the incoming photon leg, while their sum represents a
meaningful physical observable.
In the $\overline{\rm MS}$ scheme with $\xi=1$, the resolved-photon 
contribution clearly dominates for a minimum-$p_T$ cut as low as 2~GeV.
This offers the opportunity to probe the photon PDF's.
In particular, $F_g^\gamma$ is only feebly constrained experimentally.
In the forward direction, at $y_{\rm lab}>1$, it makes up more than 50\% of 
the cross section (see Fig.~\ref{fig:one}d).
Thus, a dedicated experimental study in the forward direction could help to 
pin down $F_g^\gamma$ and to distinguish between the various available
photon PDF sets.
Figure~\ref{fig:one}d also shows the predictions for the ACFGP\cite{acfgp}
set with massless charm quark and for the updated GS\cite{gs} set.
Unfortunately, the minimum-$p_T$ cut in the data is still too low to allow for
a sufficiently precise theoretical description within the QCD-improved parton 
model.

\begin{figure}[t]
\begin{tabular}{ll}
\parbox{7.3cm}{
\epsfig{file=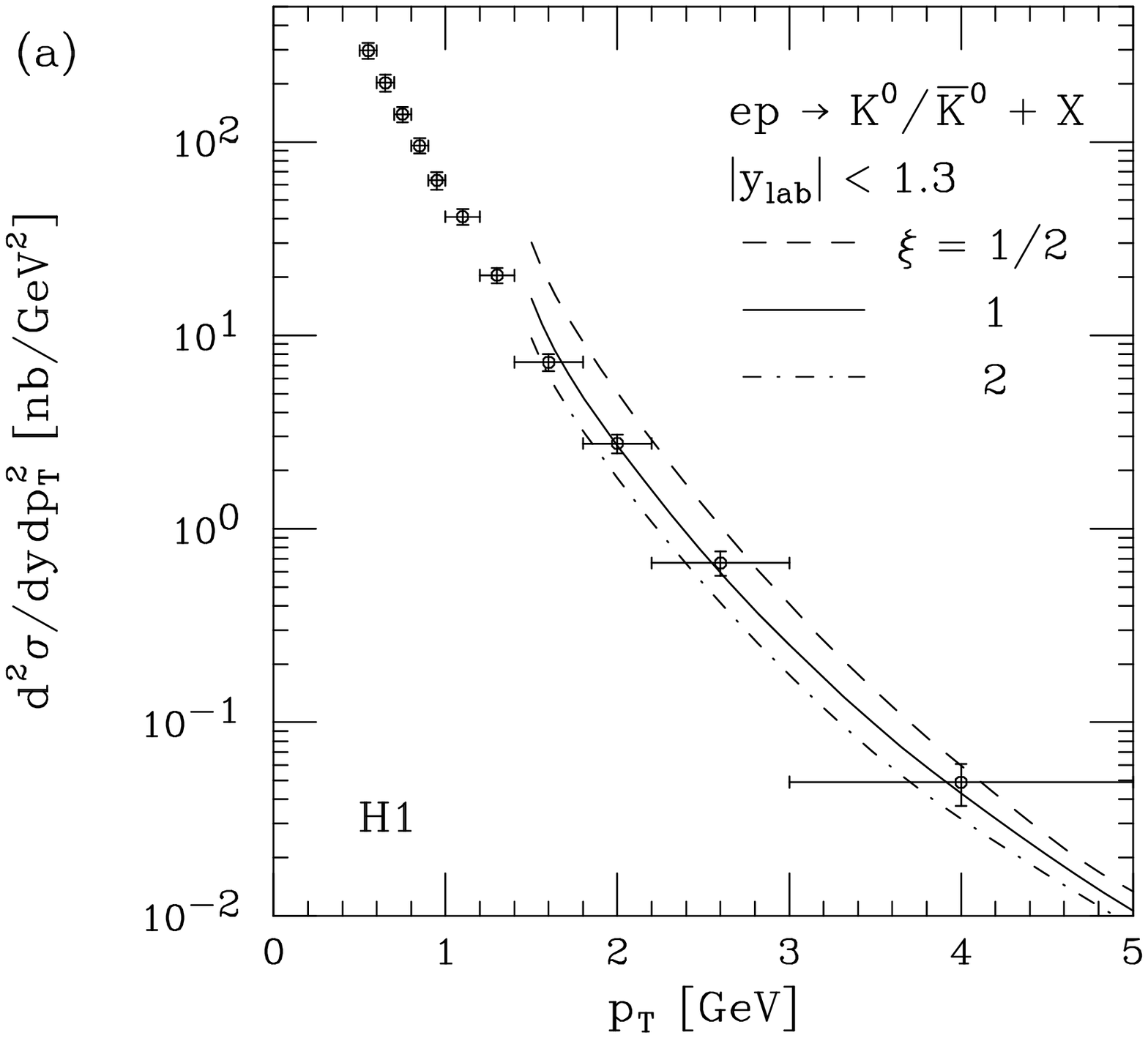,width=7.3cm}
}
&
\parbox{7.3cm}{
\epsfig{file=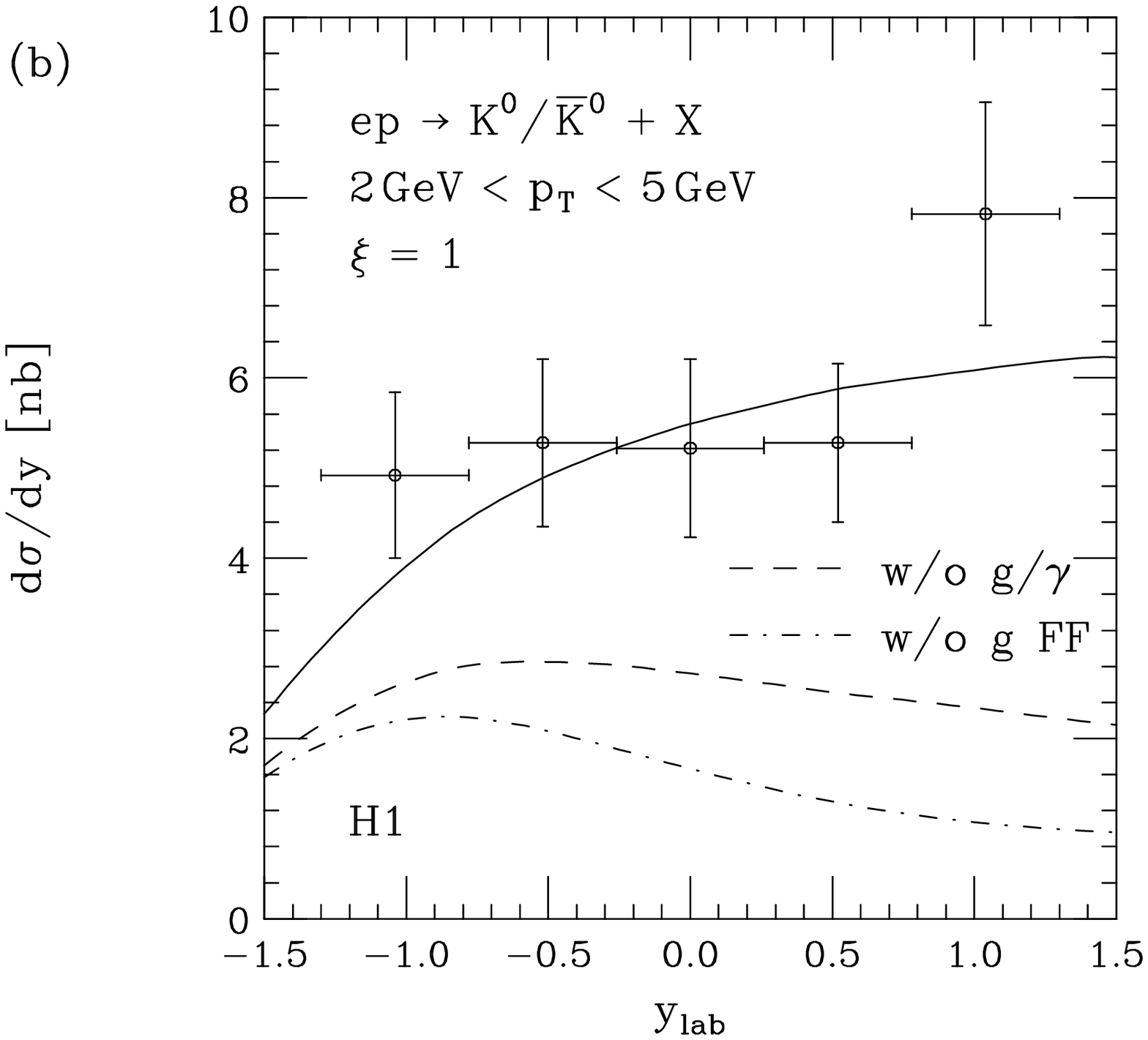,width=7.3cm}
}
\end{tabular}
\fcaption{Comparison of H1\cite{adl,joh} data on $ep\to K^0/\overline{K}^0+X$
via photoproduction with NLO predictions.}
\label{fig:two}
\end{figure}

The H1\cite{adl,joh} data on $ep\to K^0/\bar K^0+X$ via photoproduction are in 
good agreement with the corresponding NLO predictions (see
Fig.~\ref{fig:two}).
In Fig.~\ref{fig:two}b, the vertical error bars include the statistical errors
quoted in Table~4.11 of 
Ref.~11 and an overall systematic uncertainty
of\cite{joh} 10\% added in quadrature.
Strictly speaking, the minimum-$p_T$ cut of 2~GeV is too low for the
parton-model prediction to be usefully precise.
Nevertheless, Fig.~\ref{fig:two}b nicely exposes the potential of the
$y_{\rm lab}$ distribution to probe $F_g^\gamma$ and $D_g^{K^0}$.
In view of the general circumstance that the gluon FF's are less tightly
constrained by the $e^+e^-$ data than the quark FF's and the particular
assumption underlying Eq.~(\ref{eq:dg}), an independent experimental
determination of $D_g^{K^0}$ would be especially desirable.

\boldmath
\section{Comparison with data of $p\bar p$ scattering}
\unboldmath

\begin{figure}[t]
\begin{tabular}{ll}
\parbox{7.3cm}{
\epsfig{file=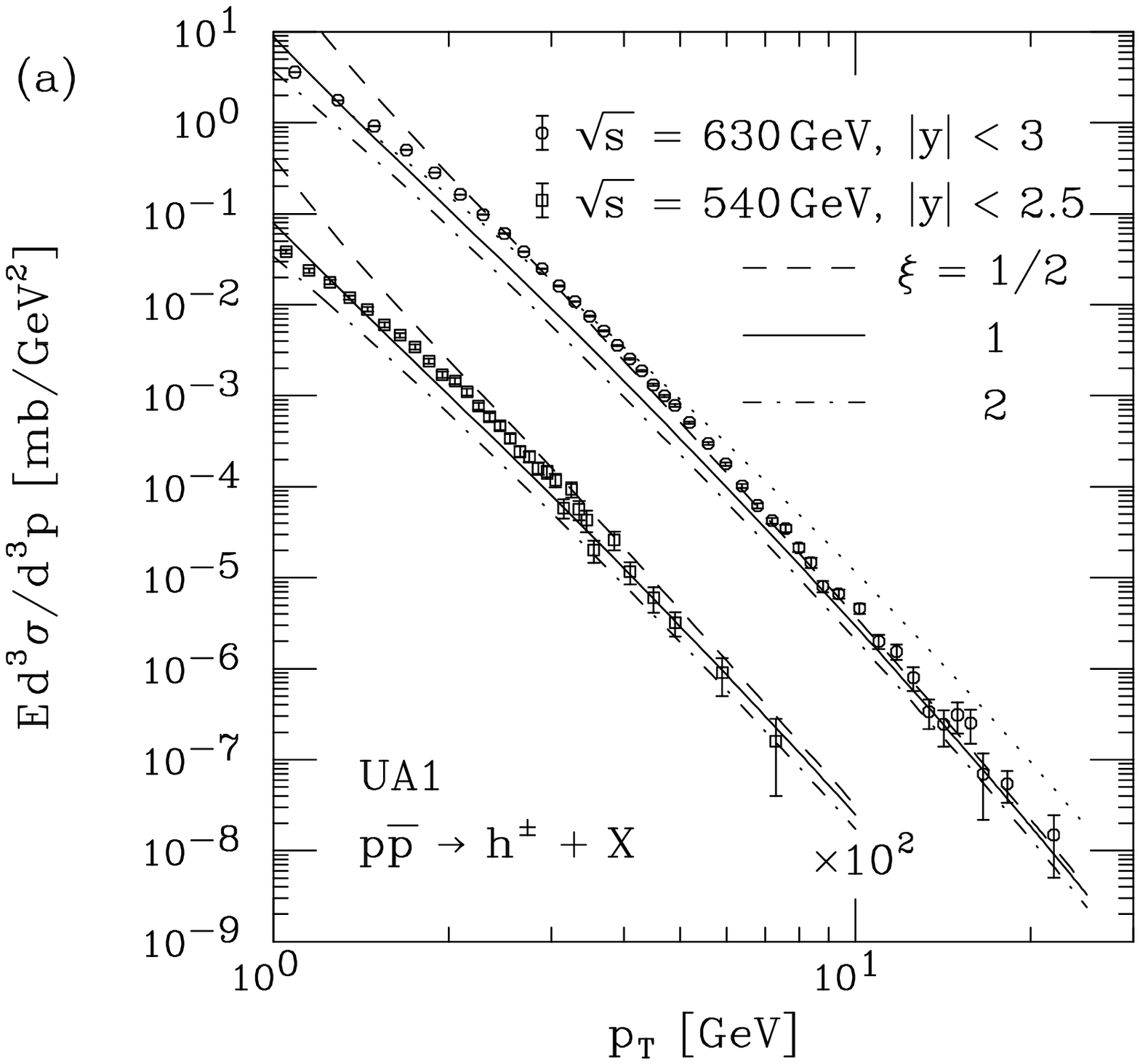,width=7.3cm}
}
&
\parbox{7.3cm}{
\epsfig{file=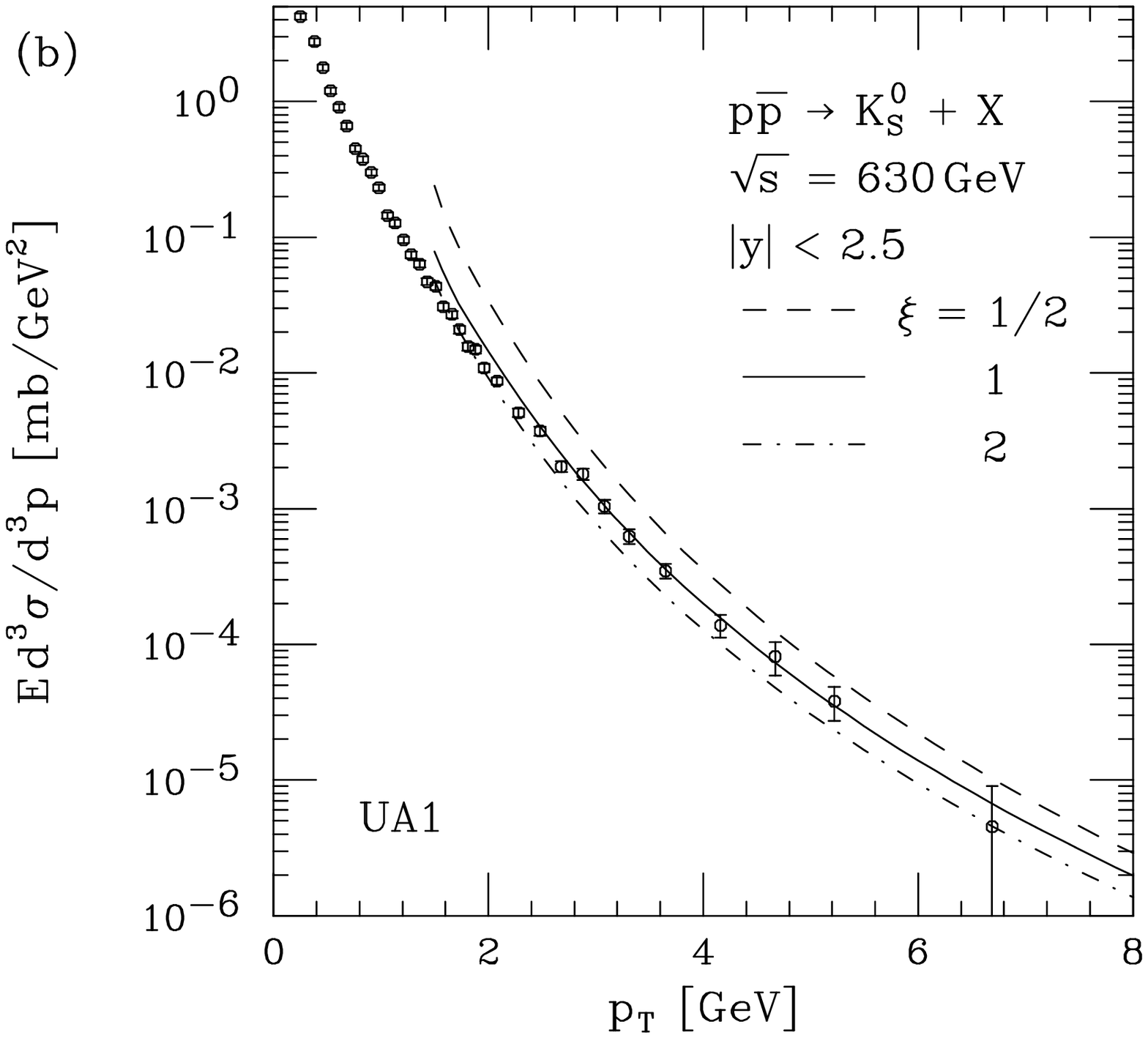,width=7.3cm}
}
\end{tabular}
\fcaption{Comparison of UA1 data on\cite{boc1} $p\bar p\to h^\pm+X$
and\cite{boc2} $p\bar p\to K_S^0+X$ with NLO predictions.}
\label{fig:three}
\end{figure}

The high-statistics data on $p\bar p\to h^\pm+X$ and $p\bar p\to K_S^0+X$
recently published by the UA1 Collaboration\cite{boc1,boc2} offer yet another
opportunity to test the universality of the FF's predicted by the
factorization theorem in a nontrivial way (see Fig.~\ref{fig:three}).
At low $p_T$, the bulks of the cross sections are due to the gluon FF's, which 
makes $p\bar p$ scattering complementary to $e^+e^-$ annihilation.\cite{bkk3}
The charged-hadron data are spread over a wide range in $p_T$, way up to
$p_T=25$~GeV, and thus carry intrinsic information on the scaling violation of
fragmentation.
This is nicely illustrated by the dotted line in Fig.~\ref{fig:three}a, which
emerges from the upper solid line by suspending the AP evolution of the FF's
and evaluating them at the fixed scale $\mu_0=\sqrt2$~GeV instead.
This leads to a significant increase at high $p_T$, by more than a factor of 5
at $p_T=25$~GeV.
The experimental data clearly favour the scaling violation encoded in the
solid line.

\section{Conclusions}

The comparative study of inclusive single-hadron production in $e^+e^-$, $ep$, 
and $p\bar p$ collisions allows for a meaningful quantitative test of the 
QCD-improved parton model and, in particular, of the scaling violation and
universality of fragmentation.
Furthermore, photoproduction experiments at HERA provide useful information on
the interplay of the direct- and resolved-photon mechanisms.
Extensions of these measurements to higher values of $p_T$ and $y_{\rm lab}$
would render an independent determination of $\alpha_s$ possible and
considerably improve our knowledge of $F_g^\gamma$.

\bigskip
\centerline{\bf ACKNOWLEDGMENTS}
\smallskip\noindent
I would like to thank Janko Binnewies and Gustav Kramer for their
collaboration on the work presented here.


\begin{thebibliography}{99}

\bibitem{gri}
V.N. Gribov and L.N. Lipatov,
{\it Yad.\ Fiz.}\ {\bf15} (1972) 781
[{\it Sov.\ J. Nucl.\ Phys.}\ {\bf15} (1972) 438];
G. Altarelli and G. Parisi,
{\it Nucl.\ Phys.}\ {\bf B126} (1977) 298;
Yu.L. Dokshitser,
{\it Zh.\ Eksp.\ Teor.\ Fiz.}\ {\bf73} (1977) 1216
[{\it Sov.\ Phys.\ JETP} {\bf46} (1977) 641].

\bibitem{bar}
V. Barger, T. Gottschalk, and R.J.N. Phillips,
{\it Phys.\ Lett.}\ {\bf70B} (1977) 51;
R. Baier, J. Engels, and B. Petersson,
{\it Z. Phys.}\ {\bf C2} (1979) 265;
M. Anselmino, P. Kroll, and E. Leader,
{\it Z. Phys.}\ {\bf C18} (1983) 307.

\bibitem{seh}
L.M. Sehgal and P.M. Zerwas,
{\it Nucl.\ Phys.}\ {\bf B108} (1976) 483;
C. Peterson, D. Schlatter, I. Schmitt, and P.M. Zerwas,
{\it Phys.\ Rev.}\ {\bf D27} (1983) 105.

\bibitem{chi}
P. Chiappetta, M. Greco, J.-Ph.\ Guillet, S. Rolli, and M. Werlen,
{\it Nucl.\ Phys.}\ {\bf B412} (1994) 3.

\bibitem{gre}
M. Greco and S. Rolli,
{\it Z. Phys.}\ {\bf C60} (1993) 169;
{\it Phys.\ Rev.}\ {\bf D52} (1995) 3853;
M. Greco, S. Rolli, and A. Vicini,
{\it Z. Phys.}\ {\bf C65} (1995) 277.

\bibitem{bkk1}
J. Binnewies, B.A. Kniehl, and G. Kramer,
{\it Z.\ Phys.}\ {\bf C65} (1995) 471.

\bibitem{bkk2}
J. Binnewies, B.A. Kniehl, and G. Kramer,
{\it Phys.\ Rev.}\ {\bf D52} (1995) 4947.

\bibitem{bkk3}
J. Binnewies, B.A. Kniehl, and G. Kramer,
{\it Phys.\ Rev.}\ {\bf D53} (1996) 3573.

\bibitem{abt}
I. Abt {\it et al.}\ (H1 Collaboration),
{\it Phys.\ Lett.}\ {\bf B328} (1994) 176;
M. Erdmann, private communication.

\bibitem{adl}
C. Adloff {\it et al.}\ (H1 Collaboration),
Report Nos.\ DESY~97--095 and hep--ex/9705018 (May 1997),
{\it Z. Phys.}\ {\bf C} (in press).

\bibitem{joh}
K. Johannsen,
Ph.D. thesis, University of Hamburg, Internal Report No.\ DESY~FH1--96--01
(June 1996).

\bibitem{der}
M. Derrick {\it et al.}\ (ZEUS Collaboration),
{\it Z. Phys.}\ {\bf C67} (1995) 227.

\bibitem{boc1}
G. Arnison {\it et al.}\ (UA1 Collaboration),
{\it Phys.\ Lett.}\ {\bf118B} (1982) 167;
G. Bocquet {\it et al.}\ (UA1 Collaboration),
{\it Phys.\ Lett.}\ {\bf B366} (1996) 434.

\bibitem{boc2}
G. Bocquet {\it et al.}\ (UA1 Collaboration),
{\it Phys.\ Lett.}\ {\bf B366} (1996) 441.

\bibitem{ell}
G. Altarelli, R.K. Ellis, G. Martinelli, and S.-Y. Pi,
{\it Nucl.\ Phys.}\ {\bf B160} (1979) 301;
R. Baier and K. Fey,
{\it Z. Phys.}\ {\bf C2} (1979) 339.

\bibitem{bkk4}
J. Binnewies, B.A. Kniehl, and G. Kramer,
Report Nos.\ DESY 97--012, MPI/PhT/97--009, and hep--ph/9702406
(February 1997), {\it Z. Phys.}\ {\bf C} (in press).

\bibitem{cow}
G.D. Cowan,
in {\it Proceedings of the XXVII International Conference on High Energy
Physics}, 20--27 July 1994, Glasgow, Scottland, UK, edited by P.J. Bussey and
I.G. Knowles (IOP, Bristol, 1995), p.~883;
D. Buskulic {\it et al.}\ (ALEPH Collaboration),
{\it Phys.\ Lett.}\ {\bf B357} (1995) 487; 
{\bf B364} (1995) 247 (E);
C. Padilla Aranda,
Ph.D. thesis, University of Barcelona (September 1995).

\bibitem{act}
P.D. Acton {\it et al.}\ (OPAL Collaboration),
{\it Z. Phys.}\ {\bf C58} (1993) 387.

\bibitem{rij}
P.J. Rijken and W.L. van Neerven,
{\it Phys.\ Lett.}\ {\bf B386} (1996) 422;
{\it Nucl.\ Phys.}\ {\bf B487} (1997) 233;
J. Binnewies,
Report Nos.\ DESY 97--128 and hep--ph/9707269 (July 1997).

\bibitem{aih}
H. Aihara {\it et al.}\ (TPC/Two Gamma Collaboration),
Report No.\ LBL--23737 (March 1988);
{\it Phys.\ Rev.\ Lett}.\ {\bf61} (1988) 1263.

\bibitem{ake}
R. Akers {\it et al.}\ (OPAL Collaboration),
{\it Z. Phys.}\ {\bf C63} (1994) 181;
D. Buskulic {\it et al.}\ (ALEPH Collaboration),
{\it Z. Phys.}\ {\bf C66} (1995) 355.

\bibitem{cla}
R. Clare {\it et al.}\ (LEP Electroweak Working Group),
Report No.\ LEPEWWG/97--02 (August 1997).

\bibitem{sch}
H. Schellman {\it et al.}\ (MARK II Collaboration),
{\it Phys.\ Rev.}\ {\bf D31} (1985) R3013.

\bibitem{dbu}
D. Buskulic {\it et al.}\ (ALEPH Collaboration),
{\it Z. Phys.}\ {\bf C64} (1994) 361.

\bibitem{wei}
E.J. Williams,
{\it Proc.\ Roy.\ Soc.\ London} {\bf A139} (1933) 163;
C.F. v.\ Weizs\"acker,
{\it Z. Phys.}\ {\bf88} (1934) 612.

\bibitem{ave}
F. Aversa, P. Chiappetta, M. Greco, and J.Ph.\ Guillet,
{\it Phys.\ Lett.}\ {\bf B210} (1988) 225; {\bf B211} (1988) 465;
{\it Nucl.\ Phys.}\ {\bf B327} (1989) 105.

\bibitem{kni}
B.A. Kniehl and G. Kramer,
{\it Z. Phys.}\ {\bf C62} (1994) 53.

\bibitem{aur}
P. Aurenche, R. Baier, A. Douiri, M. Fontannaz, and D. Schiff,
{\it Nucl.\ Phys.}\ {\bf B286} (1987) 553.

\bibitem{gor}
L.E. Gordon,
{\it Phys.\ Rev.}\ {\bf D50} (1994) 6753.

\bibitem{lai}
H.L. Lai, J. Huston, S. Kuhlmann, F. Olness, J. Owens, D. Soper, W.K. Tung,
and H. Weerts,
{\it Phys.\ Rev.}\ {\bf D55} (1997) 1280.

\bibitem{grv}
M. Gl\"uck, E. Reya, and A. Vogt,
{\it Phys.\ Rev.}\ {\bf D46} (1992) 1973.

\bibitem{acfgp}
P. Aurenche, P. Chiappetta, M. Fontannaz, J.P. Guillet, and E. Pilon,
{\it Z. Phys.}\ {\bf C56} (1992) 589.

\bibitem{gs} L.E. Gordon and J.K. Storrow,
{\it Nucl.\ Phys.}\ {\bf B489} (1997) 405.

\end{thebibliography}
\end{document}